\documentclass[10pt]{article}

\usepackage[a4paper,margin=2cm]{geometry}
\usepackage{amsmath,amssymb}
\usepackage{graphicx}
\usepackage{hyperref}
\usepackage{authblk}
\usepackage{siunitx}

\title{\textbf{Confocal Subsurface Backscattering Microscopy for Optical Identification of Nanoscale Threading Dislocations in SiC Substrates}}

\author{
Russel Cruz Sevilla$^{12,\dagger}$,
Hsiu-Ming Hsu$^{12,\dagger}$,
Irwan Saleh Kurniawan$^{12,\dagger}$,
Ruth Jeane Soebroto$^{12}$,
Hsiu-Ying Huang$^{12}$,
Jia-Ren Wu$^{23}$,
Chia-Shain Chuang$^{23}$,
Chao-Ming Fu$^{2}$,
Sheng Hsiung Chang$^{4}$,
Wen-Chung Li$^{5,*}$,
Chi-Tsu Yuan$^{12,*}$
}

\date{
\small
$^{1}$Department of Physics, Chung Yuan Christian University, Taoyuan, Taiwan\\
$^{2}$Research Center for Semiconductor Materials and Advanced Optics, Chung Yuan Christian University, Taoyuan, Taiwan\\
$^{3}$Department of Electronic Engineering, Chung Yuan Christian University, Taoyuan, Taiwan\\
$^{4}$Department of Optics and Photonics, National Central University, Taoyuan, Taiwan\\
$^{5}$WAFER WORKS, Taoyuan, Taiwan\\[0.5em]
$^{\dagger}$These authors contributed equally to this work.\\
$^{*}$Corresponding authors: ctyuan@cycu.edu.tw; wenchunglee@latentek.com.tw
}

\begin{document}

\twocolumn[
\maketitle

\begin{center}
\textbf{\large Abstract}
\end{center}

\noindent
High-density threading dislocations (TDs) in SiC wafers facilitate reverse leakage and degradation, yet commercial defect inspection systems based on surface-profiling and PL dark-contrast miss nanoscale TDs because they lack resolvable surface signatures and band-edge PL is uniformly quenched by background dopants or compensating defects. Here, we develop confocal subsurface backscattering microscopy (CSBM) to nondestructively detect TDs, based on the synergy of confocal-filtering–induced dark-field configuration and strain-induced photoelastic mechanism. By simultaneously suppressing specular reflection while enhancing optical scattering from TD-induced refractive-index perturbation, CSBM enables high-contrast, high-resolution TD imaging. Moreover, TD types can be distinguished by their distinct photoelastic scattering patterns. Our work establishes a simple but effective optical approach for direct TD identification that is more tolerant of surface imperfections, providing a practical route toward industrial in-line inspection.
\vspace{1.5em}
]

\begin{figure*}[h]
    \centering
    \includegraphics[width=\textwidth]{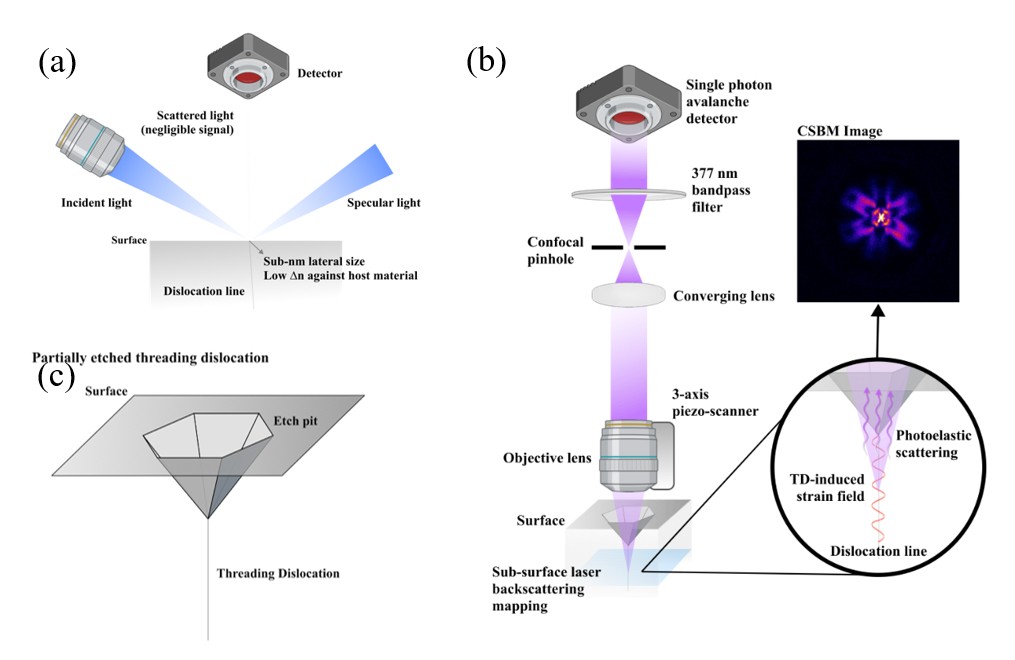}
    \caption{\textbf{Core concept of CSBM for nanoscale TD detection.} (a) Primary limitation of conventional scattering-based optical detection for TDs embedded in SiC matrix in the Rayleigh-scattering regime. (b) Schematic of CSBM, which combines confocal axial filtering to form a quasi–dark-field configuration and strain-induced photoelastic scattering. (c) Schematic of partially etched TDs used as surface markers for dislocation localization and subsequent subsurface mapping of dislocation lines.}
    \label{fig:placeholder}
\end{figure*}

\begin{figure*}[h]
    \centering
    \includegraphics[width=\textwidth]{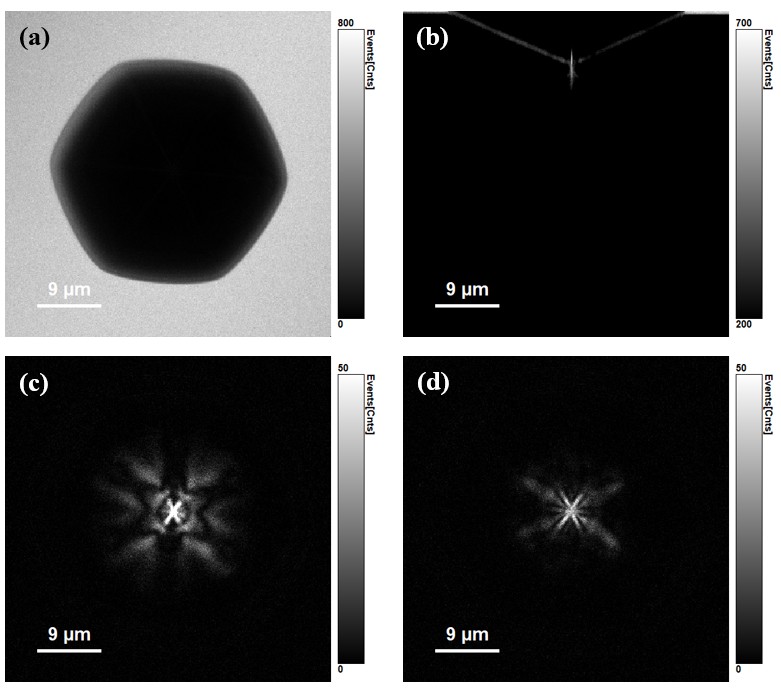}
    \caption{\textbf{Confocal backscattering microscopy of a partially etched TSD}. (a) Surface backscattering mapping image and (b) cross-sectional backscattering mapping image of the selected etched TDs. (c, d) Subsurface confocal backscattering maps of the dislocation line acquired at two representative focal depths, revealing localized bright scattering features associated with the TSD.}
    \label{fig:placeholder}
\end{figure*}

\begin{figure*}[h]
    \centering
    \includegraphics[width=\textwidth]{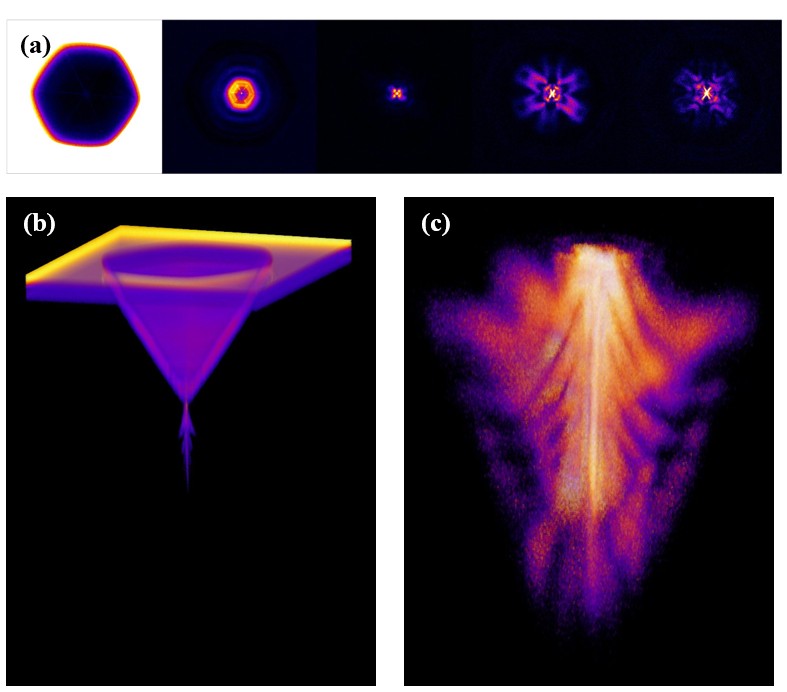}
    \caption{\textbf{2D-stack mapping and 3D reconstruction image of a TSD}(a) 2D-stack mapping of a TSD acquired from the surface to deep subsurface regions (depth increases from left to right). (b) Depth-corrected 3D reconstructed image of the TSD, including both the surface dislocation etch pit and the underlying dislocation line. (c) 3D subsurface backscattering images of the dislocation lines cropped from whole 3D image.}
    \label{fig:placeholder}
\end{figure*}

\begin{figure*}[h]
    \centering
    \includegraphics[width=\textwidth]{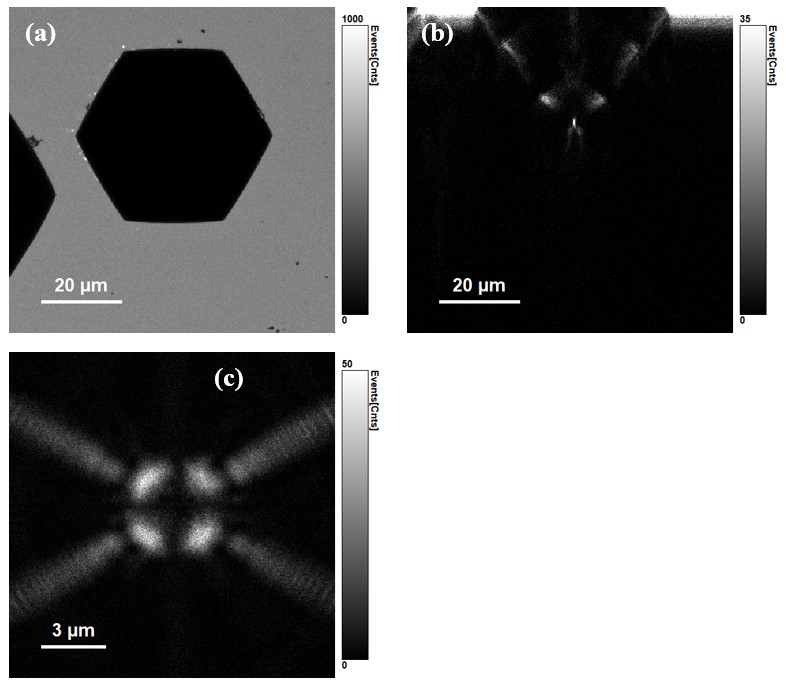}
    \caption{\textbf{Identification and subsurface backscattering pattern of a TMD}. (a) confocal surface laser backscattering image showing the surface etch pits with a comparatively large morphology. (b) Cross-sectional laser backscattering mapping beneath the selected etch pit. (c) CSBM image of the TMDs.}
    \label{fig:placeholder}
\end{figure*}

\section*{Introduction}
Threading dislocations (TDs) are among the most abundant extended defects in SiC substrates and epitaxial layers. They can facilitate leakage and act as nucleation sites for other extended defects \cite{li2022}. TDs originating in the substrate may propagate into the epitaxial layers and terminate at the surface, thereby intersecting device active regions and interacting with charge carriers along the conduction path\cite{chen2025}. As a result, TDs play a critical role in the performance and reliability of SiC-based power devices \cite{fujiwara2012,tsuji2010,lai2025,cheng2025}. Nevertheless, despite their significance, nondestructive detection of TDs remains challenging because of their extremely small physical core size (on the order of a few nanometers) and the absence of a dedicated, well-resolved luminescence signature that can be directly exploited for imaging. This limitation is particularly severe in highly n-doped or high-purity semi-insulating (HPSI) SiC substrates, where either high dopant concentrations or high densities of compensating defect centers further obscure TD-related optical signatures \cite{kurniawan2026, kato2025, harada2025}.

Commercial surface defect-inspection tools (e.g., KLA systems) that are widely used in the semiconductor industry cannot effectively detect nanoscale TDs in SiC substrates \cite{chung2023, yu2025}. TDs lack optically resolvable surface signatures, limiting surface-profiling–based optical modes. Moreover, in heavily n-doped or defective HPSI SiC substrates, abundant impurities and compensating defect centers can universally quench the band-edge PL, thereby erasing PL dark-contrast signatures. As a result, optical detection and identification of individual nanoscale TDs in SiC substrates remain challenging. To date, TDs have been mostly observed in PL imaging primarily for high-quality low-doped SiC epilayers with long carrier lifetimes, where defect-induced PL contrast is preserved \cite{feng2011, yang2024,tanuma2018}.

Laser scattering is a widely used optical approach for defect inspection, particularly for detecting surface particles and pits \cite{quan2000}. To improve signal-to-noise ratio, scattering-based inspection is often performed in a dark-field configuration to suppress specular reflection and reduce background \cite{dong2022}. However, as defect dimensions shrink to the nanoscale, the scattered signal decreases rapidly: in the Rayleigh regime, the scattering cross-section scales approximately as $a^6$ (where a is the particle diameter), causing the scattered intensity to fall quickly toward or below the noise floor \cite{zhu2022}. Consequently, even under dark-field conditions, conventional scattering-based inspection often cannot provide sufficient contrast to visualize nanoscale defects, especially TDs, whose physical core diameters are only a few nanometers. The challenge becomes even more severe for nanoscale TDs buried within the same host matrix, where the effective refractive-index contrast is weak and the scattered signal is further reduced.

Synchrotron X-ray topography (SXRT) is regarded as a state-of-the-art, nondestructive method for detecting TDs in SiC because it images strain- and lattice-tilt–induced diffraction contrast rather than relying on surface morphology or band-edge PL \cite{chen2025, shinagawa2020}. In a small-angle (grazing-incidence) geometry, the incident beam provides enhanced near-surface sensitivity to the local atomic displacement field around TDs, such that dislocations appear as characteristic features, which can be further analyzed to infer dislocation character and Burger’s vector. However, this near-surface sensitivity also makes the contrast susceptible to disturbances from surface particles, roughness, and surface imperfection \cite{black2006}. Moreover, SXRT requires access to a synchrotron beamline, careful diffraction alignment and geometry selection, and nontrivial contrast interpretation, limiting throughput and making it impractical for routine in-line defect inspection in industry \cite{michael2003}.

In this work, we develop confocal subsurface backscattering microscopy (CSBM), a purely scattering-based optical technique that requires neither polarization analysis nor phase-contrast optics, for the nondestructive detection of nanoscale TDs. The technique exploits a confocal-enabled, pseudo–dark-field configuration in combination with a strain-induced photoelastic scattering mechanism. By simultaneously suppressing specular-reflection background and enhancing backscattering from TD-induced refractive-index gradients, CSBM enables high-contrast, high-resolution visualization of TDs, even within highly defective high-purity semi-insulating (HPSI) SiC substrates. In addition, TDs can exhibit characteristic photoelastic backscattering signatures, providing a potential basis for pattern-based TD-type discrimination. Owing to its simplicity, robustness, and tolerance to surface imperfections, CSBM offers a practical and scalable approach for high-throughput, in-line industrial inspection of subsurface dislocations.

\section*{Results}
\noindent \textbf{CSBM technique for nanoscale TD detection}
In the Rayleigh scattering regime, the intensity of scattered light is highly dependent on both the physical dimensions of the scatterer and the refractive-index contrast relative to the host material. For nanoscale TDs embedded within a SiC matrix, the scattering signal produced by the dislocation core itself is remarkably weak. As a result, these defects typically fail to be detected in conventional oblique wide-field dark-field imaging or standard confocal backscattering microscopy. This intrinsic limitation renders purely scattering-based detection highly challenging, as schematically illustrated in Fig. 1a.

As a result, improving TD detectability requires strategies that both enhance TD-induced scattering and suppress the dominant specular-reflection background. Most prior approaches have focused primarily on specular-background rejection, for example through annular illumination with complementary detection and cross-polarized illumination/detection. However, background suppression alone often remains insufficient to achieve a practical signal-to-noise ratio, because the intrinsic scattering from nanoscale TD strain fields can be exceedingly weak. To address this challenge, our core concept combines a confocal axial-filtering–enabled quasi–dark-field geometry with a strain-induced photoelastic mechanism around TDs, as illustrated in Fig. 1b.

Our CSBM requires only shifting the focal plane from the surface to an optimized subsurface depth: as the focus is gradually moved into the bulk, the specular surface reflection becomes out of focus and is efficiently rejected by the confocal pinhole, thus converting a bright-field configuration into a quasi–dark-field mode. In addition, the TD-associated strain field produces a localized refractive-index perturbation through the photoelastic effect, which would selectively enhance the backscattered light from the dislocation region.

To test this hypothesis, we first carried out defect-selective etching on a HPSI SiC substrate to create partially revealed TDs. In this case, each TD consists of a surface dislocation etch pit connected to an underlying dislocation line, providing a well-defined model system for validating our proposed technique, as illustrated in Fig. 1c. Using the etch-pit positions as the markers, we can readily relocate the target TDs for subsequent surface and subsurface confocal backscattering mapping and then correlate the measured scattering signatures with the corresponding etch-pit morphologies (and thus TD types). 

Figures 2a and 2b show the surface and cross-sectional laser backscattering mapping images of a representative etched pit. In the surface backscattering image, a bright background with a distinct dark hexagonal feature is observed, corresponding to the dislocation etch pit. The cross-sectional backscattering map provides its depth profile and serves as the guidance for TD type identification. The selected pit exhibits a large hexagonal lateral morphology together with a pronounced V-shaped depth profile, which are characteristic features of threading screw dislocations (TSDs) \cite{holub2025}.

Figures 2c and 2d display the confocal subsurface backscattering maps of the dislocation lines acquired at different focal depths (depths at 36 and 42.5 microns below the surface). As the focal plane is moved into the bulk, the background gradually transitions from a bright-field–like appearance to a dark-field–like image, consistent with confocal axial filtering that suppresses out-of-focus scattering. At an optimized subsurface depth (e.g. 36 microns), a pronounced bright feature emerges at the lateral position of the TSD core. This bright core is surrounded by a distinctive star-shaped, radially lobed scattering pattern, which remains observable over a range of focal depths. The full depth-dependent evolution of the backscattering response of partially etched TSDs, from the surface to approximately \SI{65.5}{\micro\meter} into the bulk, was recorded and is provided in the Supplementary Video.

This behavior indicates that the backscattered signal is not governed solely by simple geometric scattering from a spatially localized TD core. Instead, the observed multi-lobed radiation pattern is most plausibly explained by a strain-induced photoelastic mechanism associated with the TD \cite{chen2019, fukuzawa2020}. Specifically, the long-range elastic strain field surrounding a TSD perturbs the local permittivity through the photoelastic effect, producing spatially varying refractive-index gradients in the vicinity of the dislocation line. These strain-induced index variations effectively form an extended, anisotropic scattering center that redistributes the incident optical field into preferred directions, giving rise to the characteristic star-like subsurface backscattering signature.

It should be noted that, although an ideal TSD is dominated by an azimuthal shear strain field around the dislocation line, this strain distribution can still induce spatially varying refractive-index perturbations through the photoelastic effect. The resulting index modulation, which can be further enhanced by core defect decoration, can act as an extended scattering potential. Consequently, a TSD can produce a long-range, anisotropic, and patterned subsurface backscattering signature rather than a purely localized, point-like scattering spot.

\noindent \textbf{2D-stack mapping and 3D image reconstruction of a TSD using CSBM}
To comprehensively investigate the evolution of strain-field–induced photoelastic backscattering, we performed depth-resolved 2D-stack mapping over a range of approximately 65.5 µm, starting from the surface dislocation etch pits and extending into the deep subsurface region, as shown in Fig. 3a. When the focal plane is positioned at the surface, the dislocation etch pits are clearly resolved and are dominated by geometrically induced scattering from the etched surface morphology. As the focal plane is translated away from the surface and into the bulk, the surface-related scattering is progressively suppressed by confocal axial filtering. Subsequently, the strain-induced backscattering associated with TSDs becomes dominant, giving rise to symmetric, multi-lobed, star-like scattering patterns with high contrast.

After acquiring depth-resolved 2D backscattering image stacks, the data were reconstructed to generate 3D backscattering volumes with depth correction to compensate for spherical aberration arising from refractive-index mismatch, as shown in Fig. 3b. The resulting 3D reconstruction simultaneously resolves the surface dislocation etch pit and the underlying dislocation line, enabling direct visualization of their spatial continuity from the surface into the bulk. This 3D continuity provides a direct basis for correlating TD morphology/type with its subsurface strain-induced, photoelastic backscattering signatures. To emphasize the dislocation line contribution, the reconstructed volume was further cropped to isolate the dislocation-line region, revealing a characteristic fishbone-like, anisotropic scattering pattern, as shown in Fig. 3c.

Notably, our CSBM technique provides a simple but effective approach for directly visualizing individual nanoscale TDs in HPSI-SiC substrates without any image processing or computational algorithm. In addition, this sensitivity enhancement is achieved without modifying the existing microscope hardware or added any optical components. To quantify the image quality, we evaluated the contrast-to-background ratio (CBR) at the bright central spot for the TSDs in the CSBM image, defined as $ \mathrm{CBR} = \frac{I_{TD} - I_{B}}{I_{B}} $ , where $I_{TD}$,$I_B$ denote the mean intensities of the bright spot close to TD cores and the background dark region, respectively. Using this metric, we obtain a high CBR value, reaching 206 (supporting information, S1). In addition, the spatial resolution, estimated from the full width at half maximum (FWHM) of the central brightest spot, is on the order of a few micrometers, indicating high spatial resolution.

Previous studies have shown that screw-type TDs, including pure TSDs and threading mixed dislocations (TMDs), are particularly detrimental to device performance \cite{tsuji2010, gao2023}. In an ideal elastic description, a pure TSD is dominated by an azimuthal shear strain field surrounding the dislocation line, resulting in a refractive-index perturbation that is approximately rotationally symmetric in the lateral plane through the photoelastic effect. In contrast, TMDs are expected to introduce additional strain components that break this symmetry and thus would distort the strain-induced scattering patterns, providing an opportunity for discrimination between screw-type TDs (TSDs and TMDs) using CSBM.

To demonstrate this capability, we first identify TMDs based on etch-pit morphology, specifically pit size and lateral asymmetry, using the surface and cross-sectional backscattering maps, as shown in Fig. 4a and 4b. Notably, the cross-sectional depth profile reveals a more complex and asymmetric geometry with two distinct sidewall inclinations, which is consistent with the mixed-character of TMDs. After establishing the TMD assignment, we apply CSBM to the same TMDs to acquire the corresponding subsurface backscattering map, as shown in Fig. 4c.

In contrast to TSDs, the CSBM images of the TMDs exhibit a noticeably reduced total scattering intensity and a less pronounced central spot, accompanied by a more extended spatial redistribution of the scattered light. Rather than being concentrated near the dislocation core, the backscattered signal from TMDs is spread into multiple laterally extended lobes with reduced symmetry, reflecting the presence of mixed screw–edge character. The edge component introduces an anisotropic strain field that, through the photoelastic effect, generates spatially nonuniform refractive-index perturbations, thereby redirecting the incident optical field into off-axis directions. As a result, TMDs act as extended, anisotropic scatterers with lower peak intensity but a wider scattering footprint compared with the more compact and centrally dominated patterns observed for pure TSDs.

It should be noted that our CSBM is a non-polarization, purely scattering-based technique. As a result, the image contrast is dominated by localized TD-associated strain fields that generate photoelastic backscattering within the confocal volume, while more broadly distributed residual stress and weak large-area inhomogeneity typically contribute a lower, slowly varying background. This selective sensitivity enables high spatial resolution and strong defect contrast, while reducing background signals from out-of-focus regions and widely distributed strain. In contrast, polarization-based techniques such as polarized light microscopy are highly sensitive to strain-induced birefringence and depolarization; therefore, residual stress and abundant defects can introduce substantial background intensity variations that may obscure subtle defect-specific contrast. 

\section*{Discussion}
\noindent \textbf{Detection sensitivity of nanoscale TDs based on CSBM}
Although CSBM enables reliable detection of individual TDs, it is important to clarify why such nanoscale defects are resolvable with this approach yet are typically missed in conventional wide-field surface inspection, even under dark-field conditions. If one considers only the atomic-scale dislocation core embedded in a homogeneous SiC host, a simple Rayleigh-type estimation for a subwavelength scattering volume with a weak refractive-index perturbation yields an extremely small scattering cross section. This arises from the steep $\sigma \propto a^6$ scaling together with the intrinsically tiny difference of refractive index between the TDs and their host (Supporting Information S2). Consequently, purely geometric scattering from the core alone is far below the detection threshold of conventional optical methods, explaining why standard wide-field or dark-field imaging fails to reveal nanoscale TDs.

In contrast, if we treat TDs as strain-assisted scatterers, a Rayleigh model modified by the photoelastic mechanism can be used to roughly estimate the sensitivity and detectivity of our CSBM technique. In the low-perturbation regime, the elastic scattering cross-section can be written as,$\sigma_R \approx \frac{8\pi}{3} k^4 a^6 \left(\frac{\Delta n}{n}\right)^2$ , where $k=\frac{2\pi}{\lambda}$, a is the effective radius of the strain-perturbed region, and $\frac{\Delta n}{n}$ is the strain-induced refractive-index change. In addition, the strain-induced index change can be approximated as $\Delta n=n^3 p\epsilon$, where n is the refractive index of 4H-SiC, p is a photoelastic coefficient, and $\epsilon$ denotes the local strain magnitude. 

For an excitation wavelength of 375 nm and a high-NA (0.9) objective, the diffraction-limited focal spot yields a photon flux density on the order of $10^{24}$ photons/s $\times cm^2$ at an incident power of 1 mW. Under this conduction, for a given p$\sim$0.1, we obtain $\frac{\Delta n}{n}\sim 10^{-4}\sim 10^{-2}$ for $\epsilon\sim 10^{-4} \sim 10^{-2}$ within the confocal voxel. Substituting these values into the equation, with the radius of a$\sim$ 50–100 nm, gives Rayleigh cross-section $\sigma_R \approx 10^{-15}\sim 10^{-17}$. In this case, the scattered photon rate can reach $10^6\sim 10^8$ photons/s. Accounting for a collection efficiency of $\sim$0.3 and an overall detection efficiency of $\sim$0.2 (considering all optics and detector), this translates into detected photon count rates of $10^4\sim 10^6$ counts/s per TD. These values largely exceed the dark-count level of our APD detectors, confirming that CSBM provides sufficient sensitivity to resolve the strain-induced scattering from individual nanoscale TDs in HPSI 4H-SiC substrates.

\noindent \textbf{Benefits of our CSBM technique}
Conventional surface laser-scattering inspection typically relies on dark-field strategies, such as grazing-angle oblique illumination, confocal complementary illumination/detection, and cross-polarized illumination/detection, to suppress the dominant specular-reflection background and improve the signal-to-noise ratio (SNR) \cite{yoo2001, hu2025, liu2011}. However, as the characteristic size of scattering features approaches the nanoscale, the scattered signal decreases rapidly and can fall below the detection noise floor, leading to severe SNR degradation and making reliable TD detection increasingly challenging. X-ray topography (XRT) is among the most widely used nondestructive techniques for imaging TDs in SiC, where contrast originates from dislocation-induced lattice distortions and the resulting diffraction perturbations \cite{kallinger2011, senzaki2024}. Nevertheless, grazing-incidence XRT often requires synchrotron radiation and specialized beamline instrumentation, which limits its practicality for high-throughput, in-line industrial defect inspection.

PL imaging is a widely used, commercially available, nondestructive approach for defect inspection in wide-bandgap semiconductors \cite{das2018}. Under above-bandgap excitation, defects that act as efficient nonradiative recombination centers locally quench the near-band-edge emission, giving rise to dark spots in PL images \cite{yang2024}. In practice, this method is most effective when the host exhibits sufficiently strong and spatially uniform band-edge PL, so that target defects produce a measurable local reduction. Unfortunately, when other defect species or impurities are spatially widespread, the band-edge PL can be uniformly quenched across the sample, substantially reducing PL-image contrast or rendering the method ineffective \cite{feng2011}. This situation is common for SiC substrates that are highly doped or highly defective, where a high density of compensating centers can suppress band-edge emission throughout the HPSI-SiC substrates.

To date, there remains a strong demand for an optical technique capable of detecting and identifying nanoscale TDs with high image contrast and spatial resolution, particularly in SiC substrates, to enable nondestructive, in-line inspection in the wide-bandgap semiconductor industry. For example, Harada et al. combined synchrotron X-ray diffraction with birefringence measurements to identify TDs \cite{harada20252}. In parallel, Hattori et al. showed that integrating phase-contrast imaging with polarization analysis can discriminate between TSDs and TEDs, although the achievable image contrast remains relatively limited \cite{hattori2025}. More recently, M. Kato et al, have demonstrated a focused light birefringence for 3D observation of dislocations in SiC but with limited image contrast and spatial resolution \cite{kato2025}. In addition, PL-based approaches have been demonstrated in which TD-related emission is activated via deep-level states, providing an alternative route for probing electrically active TDs \cite{kurniawan2026}. 

In conclusion, we have developed an optical technique that enables nondestructive detection of individual nanoscale TDs with high image contrast and spatial resolution. The method operates by synergistically combining a confocal axial-filtering–enabled quasi–dark-field configuration with the strain-induced photoelastic effect. In this scheme, the specular-reflection background is strongly suppressed, while scattering from TD-induced refractive-index perturbations is enhanced. Moreover, the resulting photoelastic scattering morphologies provide characteristic subsurface signatures that can assist in differentiating dislocation types. Our work demonstrates a simple yet effective approach for detecting and characterizing TDs in highly defective HPSI-SiC substrates, offering a promising route toward high-throughput, in-line inspection in wide-bandgap semiconductor manufacturing.

\section*{Material and Methods}
\textbf{Materials}
The HPSI-SiC substrate was provided by SICC Co. Ltd. The substrate was grown using physical vapor transport (PVT). The Potassium Hydroxide (KOH) pellets (ca. 85$\%$) were purchased from Thermo Fisher Scientific Inc.

\noindent \textbf{Etching Process}
Prior to etching, the HPSI-SiC substrate was thoroughly cleaned using ethanol, acetone, and DI water to remove organic pollutants, followed by nitrogen drying. Molten KOH was used as etchant. The substrate was placed on a ceramic crucible and inside a temperature-controlled furnace together with the KOH pellets under 510 ºC for 15 mins. After etching, the sample was immediately quenched with DI water and sonicated with ethanol, and then with acetone, to remove residuals on the surface. 

\noindent \textbf{Confocal Subsurface Mapping}
Our confocal subsurface backscattering microscopy (CSBM) is implemented based on a conventional confocal microscope (MT100, PicoQuant). A pulsed laser with a central wavelength of $\sim$375 nm is tightly focused into the subsurface region containing the target TDs using a high–numerical-aperture (NA = 0.9) objective. The lateral spatial resolution at the surface can be simply estimated as $\sim$0.25 µm from $\frac{0.61\lambda}{NA}$ , and is slightly degraded at subsurface planes due to refractive-index mismatch. The backscattered light is collected by the same objective, passed through a bandpass filter and a confocal pinhole, and detected using an avalanche single-photon counting photodiode. For confocal subsurface mapping, the objective is mounted on a three-dimensional piezo scanner, which is used to scan the focal volume within the subsurface plane and record the backscattered intensity distribution.

\end{document}